\documentclass[twocolumn,prstab,showpacs,raggedbottom]{revtex4}

\usepackage{graphicx}
\usepackage{amsmath}
\usepackage{subfigure}

\begin{document}

	\title{Reversible electron beam heating for suppression of microbunching instabilities at free-electron lasers}

	\author{Christopher Behrens$^\mathrm{1}$, Zhirong Huang$^\mathrm{2}$, and Dao Xiang$^\mathrm{2}$}
	\affiliation{$^\mathrm{1}$ Deutsches Elektronen-Synchrotron DESY, Notkestr.\,85, 22607 Hamburg, Germany\\
	\mbox{$^\mathrm{2}$  SLAC National Accelerator Laboratory, Menlo Park, CA 94025, USA}}

	\date{\today}				
	
	\begin{abstract}
	The presence of microbunching instabilities due to the compression of high-brightness electron beams at existing and future X-ray free-electron lasers (FELs) results in restrictions on the attainable lasing performance and renders beam imaging with optical transition radiation impossible. The instability can be suppressed by introducing additional energy spread, i.e., ``heating'' the electron beam, as demonstrated by the successful operation of the laser heater system at the Linac Coherent Light Source. The increased energy spread is typically tolerable for self-amplified spontaneous emission FELs but limits the effectiveness of advanced FEL schemes such as seeding. In this paper, we present a reversible electron beam heating system based on two transverse deflecting radio-frequency structures (TDSs) up and downstream of a magnetic bunch compressor chicane. The additional energy spread is introduced in the first TDS, which suppresses the microbunching instability, and then is eliminated in the second TDS. We show the feasibility of the microbunching gain suppression based on calculations and simulations including the effects of coherent synchrotron radiation. Acceptable electron beam and radio-frequency jitter are identified, and inherent options for diagnostics and on-line monitoring of the electron beam's longitudinal phase space are discussed.

	\end{abstract}


	\maketitle

	\section{Introduction}
	X-ray free-electron lasers (FELs) provide an outstanding tool for studying matter at ultrafast time and atomic length scales~\cite{LCLSnature1}, and have become a reality with the operation of the Free-Electron Laser in Hamburg (FLASH)~\cite{FLASHAnature}, the Linac Coherent Light Source (LCLS)~\cite{LCLSnature2}, and the SPring-8 Angstrom Compact Free Electron Laser (SACLA)~\cite{SACLAnature}. The required high transverse and longitudinal brightness of the X-ray FEL driving electron bunches may encounter various degradation effects due to collective effects like coherent synchrotron radiation (CSR) or microbunching instabilities (e.g., Refs.~\cite{CSR,CSR-ub,lsc-ub}), and need to be preserved and controlled. In order to suppress a microbunching instability associated with longitudinal bunch compression that deteriorates the FEL performance, the LCLS uses a laser heater system to irreversibly increase the uncorrelated energy spread within the electron bunches, i.e., the slice energy spread, to a level tolerable for operation of a self-amplified spontaneous emission FEL~\cite{LH,LCLSheater}. For future X-ray FELs that plan to use external quantum lasers (seed lasers) to seed the FEL process in order to achieve better temporal coherence and synchronization for pump-probe experiments, a smaller slice energy spread is required to leave room for the additional energy modulation imprinted by the seed laser. Thus, the amount of tolerable beam heating is more restrictive and the longitudinal phase space control becomes more critical (e.g., Refs.~\cite{NLS,flatflat}). The same strict requirement on small slice energy spreads is valid for optical klystron enhanced self-amplified spontaneous emission free-electron lasers~\cite{okly}.

	Originally designed for high-energy particle separation by radio-frequency (rf) fields~\cite{LOLA}, transverse deflecting rf structures (TDSs) are routinely used for high-resolution temporal electron beam diagnostics at present X-ray FELs (e.g., Refs.~\cite{Kick,Roehrs,Filippetto,xtcav}) and are proposed to use for novel beam manipulation methods (e.g., Refs.~\cite{psxray, exchange1, exchange2, mapping, ramp, psex}). Recently, a TDS was used to increase the slice energy spread in an echo-enabled harmonic generation FEL experiment~\cite{Xiang,Xiang2}. In this paper, we present a reversible electron beam heating system that uses two TDSs located up and downstream of a magnetic bunch compressor chicane. The additional slice energy spread is introduced in the first TDS, which suppresses the microbunching instability, and then is eliminated in the second TDS.

	The method of reversible beam heating is shown in Sec.~\ref{sec:Method} by means of linear beam optics and a corresponding matrix formalism. In Sec.~\ref{sec:Heating}, we show the feasibility of this scheme to preserve both the transverse and longitudinal brightness of the electron beam, and discuss the impact of coherent synchrotron radiation. Section~\ref{sec:uB} covers the gain suppression of microbunching instabilities by analytical calculations and numerical simulations, and in Sec.~\ref{sec:Practical} we discuss the impact of beam and rf jitter, and show inherent options for diagnosis and on-line monitoring of the electron beam's longitudinal phase space. The results and conclusions are summarized in Sec.~\ref{sec:Summary}.

	\section{Method}\label{sec:Method}
		\begin{figure*}[htb]
	\centering
	\includegraphics*[width=0.9\linewidth]{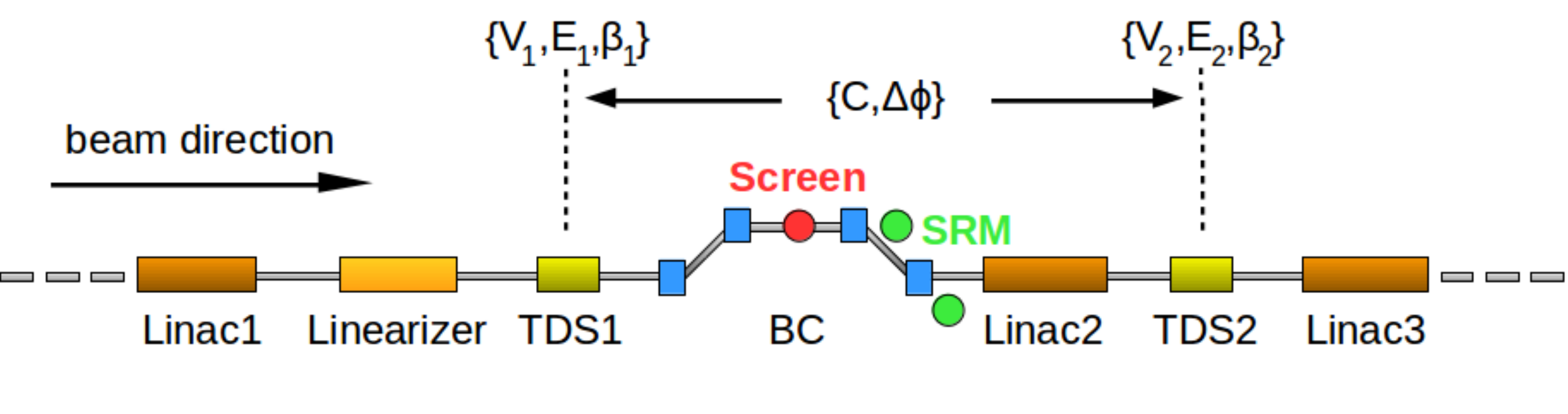}
	\caption{Layout of a reversible electron beam heater system including two transverse deflecting rf structures located up and downstream of a magnetic bunch compressor (BC) chicane, and longitudinal phase space diagnostics using screens and synchrotron radiation monitors (SRM). Parameters related to the reversible beam heater system are denoted in curly brackets.} \label{fig:Fig1_Setup}
	\end{figure*}	
	In this and the following sections, we consider a linear accelerator (linac) employing a single bunch compressor for a soft X-ray FEL, such as the proposed linac configuration for the Next Generation Light Source (NGLS) at LBNL~\cite{Corlett}. The choice of a single magnetic bunch compressor simplifies our consideration and analysis, although the concept is also applicable for typical bunch compressor arrangements with multiple stages. We note that a single bunch compressor arrangement has also been considered for the FERMI@Elettra FEL in order to minimize the impact of microbunching instabilities~\cite{Venturini}.

	The generic layout of the reversible electron beam heater system is depicted in Fig.~\ref{fig:Fig1_Setup}. It consists of linac sections providing and accelerating high-brightness electron beams, a magnetic bunch compressor chicane in order to achieve sufficient peak currents to drive the FEL process, and two transverse deflecting rf structures located up and downstream of the bunch compressor. An additional higher-harmonic rf linearizer system (Linearizer), like at the LCLS or FLASH~\cite{Lin}, can be used to achieve uniform bunch compression by means of longitudinal phase space linearization upstream of the bunch compressor. The whole system can be supplemented by dedicated longitudinal phase space diagnostics (see Sec.~\ref{sec:Practical}), and except for the two TDSs, the layout is commonly used for bunch compression at present and future X-ray FELs.

	 The principle of the reversible electron beam heater relies on the physics of TDSs arising from the Panofsky-Wenzel theorem~\cite{Panowsky,Bro}, which states that the transverse momentum gain $\Delta \vec{p}_{\perp}$ of a relativistic electron imprinted by a TDS is related to the transverse gradient of the longitudinal electric field $\nabla_{\perp}\mathcal{E}_z$ inside the TDS, and yields
	\begin{equation}
	\Delta \vec{p}_{\perp} = -i \frac{e}{\omega} \int_0^L \nabla_{\perp} \mathcal{E}_z d\tilde{z}\,,
	\end{equation} where $\omega/(2\pi)$ is the operating rf frequency, $e$ is the elementary charge, $L$ is the structure length, and $\tilde{z}$ is the longitudinal position inside the structure (not to be confused with the beamline coordinate, which is given by $s$ in the following). Operating a TDS with vertical deflection, i.e., in y-direction, near the zero-crossing rf phase $\psi=\omega/c\,z$, electrons experience transverse kicks~\cite{Kick}
	\begin{equation}
	\Delta y'=\frac{e \omega V_y}{cE}z = K_y z
	\label{eq:long}
	\end{equation} and relative energy deviations ($\delta=\Delta E/E$)~\cite{IES1,IES2}
	\begin{equation}
	\Delta \delta = K_y \frac{1}{L}\int_{0}^L{y(s) ds}=K_y \overline{y}\,,
	\label{eq:ies}
	\end{equation} where $K_y = e \omega V_y/(cE)$ is the vertical kick strength, $V_y$ is the peak deflection voltage in the TDS, $c$ is the speed of light in vacuum, $E$ is the electron energy, and $\overline{y}$ is the mean vertical position over the structure length $L$ along the beamline relative to the central axis inside the finite TDS. Here, $z$ is the internal bunch length coordinate of the electron relative to the zero-crossing rf phase. Both the additional transverse kicks and relative energy deviations are induced by the TDS operation itself and generate correlations within an electron bunch. In fact, near the zero-crossing rf phase (see Eq.~(\ref{eq:long})), the induced transverse kick correlates linearly with the internal bunch length coordinate ($z=ct$) and enables high-resolution temporal diagnostics (e.g., Refs.~\cite{Kick,Roehrs,Filippetto}), whereas the induced relative energy deviation correlates with the vertical offset inside the TDS and results in an induced relative energy spread $\Delta \sigma_{\delta} = K_y \sigma_{y}$. Here, the symbol $\sigma$ denotes the root mean square (r.m.s.) value, and $\sigma_{y}$ is the vertical r.m.s.~beam size. This additional energy spread (cf. laser heater~\cite{LH,LCLSheater}), in combination with the momentum compaction $R_{56}$ of a bunch compressor chicane, is able to smear microbunch structures, and correspondingly suppresses the associated instability as is shown in Sec.~\ref{sec:uB}. The effect of induced energy spread (``beam heating'') is generated by off-axis longitudinal electric fields, related to the principle of a TDS by the Panofsky-Wenzel theorem, and has been observed experimentally at FLASH~\cite{IES3} and the LCLS~\cite{emma}. The induced energy spread is uncorrelated in the longitudinal phase space $(z,\delta)$, but shows correlations in the phase space $(y,\delta)$, which is the reason that it can be eliminated (``beam cooling'') with a second TDS in a reversible mode as is shown in the following by two different approaches.

		\subsection{Linear beam optics}
		
		The transverse betatron motion of an electron passing through a TDS with vertical deflection (in $y$) is given by 
		\begin{equation}
			y(s)=y_0(s) + S_y(s,s_0)z
		\label{eq:motion}
		\end{equation} with the vertical shear function (e.g., Refs.~\cite{Kick,Roehrs,IES2})
		\begin{equation}
			 S_y(s,s_0) = R_{34}K_y= \sqrt{\beta_y(s)\beta_y(s_0)}\mathrm{sin}(\Delta\phi_y(s,s_0)) \frac{e \omega V_y}{cE}\,,
		\label{eq:beta}
		\end{equation} where $R_{34}$ is the angular-to-spatial element of the vertical beam transfer matrix from the TDS at $s_0$ to any position $s$, $\beta_y$ is the vertical beta function, $\Delta\phi_y$ is the vertical phase advance between $s_0$ and $s$, and $y_0$ describes the vertical beam offset independent of any TDS shearing effect. Referring to the layout depicted in Fig.~\ref{fig:Fig1_Setup} and taking bunch compression into account, the induced vertical beam offset ($\Delta y=y-y_0$) downstream of the second TDS becomes (omitting the subscript $y$ in $S_{y}$)
		\begin{align}
			\Delta y(s)= & S_1(s,s_1)z_1 + S_2(s,s_2)z_2 \nonumber \\
				= &(CS_1(s,s_1) + S_2(s,s_2))z_2
		\label{eq:sc}
		\end{align} with the bunch compression factor $C=z_1/z_2$ and the shear functions $S_{1,2}(s,s_{1,2})$ of the corresponding TDSs. Here, $ S_1(s,s_1)z_1$ describes the vertical beam offset induced by the first TDS located at $s_1$ that is independent of the second TDS. In order to cancel the spatial chirp induced by the combined TDS operation, the beam offset $\Delta y$ in Eq.~(\ref{eq:sc}) must vanish for any $z_2$. Hence, using Eq.~(\ref{eq:beta}) for $S_{1,2}$ in Eq.~(\ref{eq:sc}) and taking acceleration from $E_1$ to $E_2$ in Linac2 into account by making the replacement $\beta_y(s)\beta_y(s_1)\rightarrow\beta_y(s)\beta_y(s_1) E_1/E_2$~\cite{Kick}, we get
		\begin{align}
			&C\sqrt{\beta_{y}(s_1)}\mathrm{sin}(\Delta\phi_{y}(s,s_1))\sqrt{E_1}K_1 \nonumber \\
			&+ \sqrt{\beta_{y}(s_2)}\mathrm{sin}(\Delta\phi_{y}(s,s_2))\sqrt{E_2}K_2=0 \,,
			\label{eq:prior}
		\end{align} where $K_{1,2}$ are the vertical kick strengths of the corresponding TDSs, and $\Delta\phi_{y}(s,s_{1,2})$ describes the vertical phase advances between $s_{1,2}$ and $s$, respectively. As a consequence, the phase advance between both TDSs is $\Delta\phi_y(s_2,s_1)=\Delta\phi_{y}(s,s_1)-\Delta\phi_{y}(s,s_2)$. A general solution, valid for any position $s$ downstream of the second TDS, is only possible for a phase advance difference of
		\begin{equation}
		\Delta\phi_y(s_2,s_1)=n\cdot\pi
		\label{eq:case}
		\end{equation} with $n$ being integer, and the kick strength
		\begin{equation}
		 K_2 =\pm	C\sqrt\frac{\beta_{y}(s_1)}{\beta_{y}(s_2)}\sqrt{\frac{E_1}{E_2}}K_1\,,
		\label{eq:case1}
		\end{equation} where the sign depends on the actual phase advance, i.e., $\Delta\phi_y(s_2,s_1)=\pi + n\cdot2\pi$ for ($+$) and $\Delta\phi_y(s_2,s_1)= n\cdot2\pi$ for ($-$). The different sign of $K$ can technically be achieved by changing the rf phase in the TDS by $180^\circ$ which results in a zero-crossing rf phase with opposite slope and deflection. Besides cancelation of the induced spatial chirps, the induced energy spread of the first TDS needs to be eliminated in the second structure in order to have a fully reversible electron beam heater. Applying Eq.~(\ref{eq:ies}) similar to Eq.~(\ref{eq:sc}), the relative energy deviation downstream of the second TDS for finite structure lengths become (omitting the argument in $y(s)$ and $S(s)$)
		\begin{equation}
			\Delta \delta= K_1\overline{y_1}C\frac{ E_1}{E_2} + K_2 \overline{(y_2+S_1 z_1)}
		\label{eq:es}
		\end{equation} with the mean vertical offsets $\overline{y_1}$ and $\overline{(y_2+S_1 z_1)}$ inside the TDSs. For constant vertical offsets inside the TDSs or short structure lengths, the mean vertical offsets can be replaced by the actual offsets, i.e., $\overline{y_1} \rightarrow y_1$ and $\overline{(y_2+S_1 z_1)} \rightarrow (y_2+S_1 z_1)$. The latter describes the offset in the second TDS and involves the spatial chirp induced by the first TDS with $S_1 \sim \mathrm{sin}(\Delta\phi_y(s_2,s_1))$, which vanishes in the case of spatial chirp cancelation given by Eq.~(\ref{eq:case}). In order to cancel the relative energy spread induced by the combined TDS operation, it follows
		\begin{equation}
			 K_1\overline{y_1}C\frac{ E_1}{E_2} + K_2 \overline{y_2} =0\,.
		\label{eq:es2}
		\end{equation} The general transverse beam transport optics with the vertical phase advance condition in Eq.~(\ref{eq:case}) gives $\overline{y_2}=\pm \overline{y_1}\sqrt{\beta_{y}(s_2)/\beta_{y}(s_1)}$, and taking $\beta_{y}(s_2)\rightarrow\beta_{y}(s_2) E_1/E_2$ (see prior Eq.~(\ref{eq:prior})) into account yields exactly the same condition as in Eq.~(\ref{eq:case1}). Simultaneous spatial chirp and energy spread cancelation in the second TDS is the basic principle for reversible electron beam heating and enables local increase of slice energy spread. The additional energy spread in the bunch compressor, which is added in quadrature by the first TDS, can be controlled by the kick strength $K_1$ and the vertical beam size $\sigma_{y}(s_1)$.
		
		In the following, a complementary approach to discuss the reversible beam heating system is shown. It uses the matrix formalism for beam transport and provides an analytical way to show microbunching gain suppression and to discuss the impact of beam and rf jitter.

		\subsection{Matrix formalism}
		We adopt the beam transport matrix notation of a 6x6 matrix for $(x,x',y,y',z,\delta)$ but leaves $(x,x')$ out for simplicity, i.e., $(y,y',z,\delta)$ is used in the following. The 4x4 beam transport matrix for a vertical deflecting TDS in thin-lens approximation reads (e.g., Refs.~\cite{exchange1,IES1,IES3})
		\begin{equation}
		{\mathbf R}_T^{thin}=\begin{pmatrix}
		1 & 0 & 0  & 0 \\
		0 & 1 & K & 0 \\
		0 & 0 & 1 &0 \\
		K &0 & 0 & 1
		\end{pmatrix}.
		\label{eq:MTDS_thin}
		\end{equation} As discussed above, the main components of the given reversible heater system shown in Fig.~\ref{fig:Fig1_Setup} consist of TDS1 with the kick strength $K_1$, a bunch compressor with the momentum compaction factor $R_{56}$, and TDS2 with the kick strength $K_2$. Including the momentum compaction factor $R_{56}$ and acceleration in Linac2 ($E_1\rightarrow E_2$), the 4x4 beam matrix between the two TDSs is given by
		\begin{equation}
		{\mathbf R}_C=\begin{pmatrix}
		R_{33} & R_{34} & 0  & 0 \\
		R_{43} & R_{44} & 0 & 0 \\
		0 & 0 & 1 &R_{56} \\
		0 &0 & 0 & \frac{E_1}{E_2}
		\end{pmatrix}\,.
		\label{eq:-I}
		\end{equation}
		In order to allow the energy change in the first TDS to be compensated for in the second TDS, we require the point-to-point imaging from TDS1 to TDS2 (i.e., $R_{34}=0$), which corresponds to an equivalent vertical phase advance of $\Delta\phi_y(s_2,s_1)=n\cdot\pi$ with $n$ being integer (see Eq.~(\ref{eq:case})). Then we get the magnification factor $R_{33}=\pm\sqrt{\beta_{y}(s_2)/\beta_{y}(s_1)}$ and $R_{44}=1/R_{33}$.

		The linear accelerator section with higher-harmonic rf linearizer (Linac1 and Linearizer) upstream of the first TDS introduces an appropriate energy chirp $h$ for uniform bunch compression. Without loss of generality, we neglect acceleration between the two TDSs, i.e., we do not consider Linac2 anymore. Including Linac2 would simply result in a correction term $\sqrt{E_1/E_2}$ (cf. Eqs.~(\ref{eq:case1}) and~(\ref{eq:cond}) below) but would leave the physics unchanged. Then the entire 4x4 beam transport matrix from the beginning of TDS1 to the end of TDS2 becomes
		\begin{equation}
		\begin{pmatrix}
		R_{33} & 0 & 0  & 0 \\
		R_{43}+K_1 K_2 R_{56} & \frac{1}{R_{33}} & \frac{K_1}{R_{33}}+K_2 (1+hR_{56})  & K_2 R_{56} \\
		K_1 R_{56} & 0 & 1+hR_{56} & R_{56} \\
		{K_1}+R_{33}K_2  & 0 & 0 & \frac{1}{1+hR_{56}}
		\end{pmatrix}\,.
		\label{eq:MTDS}
		\end{equation} Cancelation of the induced spatial chirp ($\Delta y' \sim z_0$, cf. Eq.~(\ref{eq:long})) requires $R_{45}=0$ (6x6-matrix notation), i.e.,
		\begin{equation}
		K_1/R_{33}+K_2 (1+h R_{56})=0\,,
		\label{eq:cond}
		\end{equation} where $R_{45}$ describes the coupling between $y'$ and $z_0$. We note that the coupling between $\delta$ and $y_0$ (i.e., $R_{63}$ element) is nonzero in Eq.~(\ref{eq:MTDS}) because the bunch is energy-chirped after compression ($\delta\sim z\sim y_0$), which  can be removed by Linac3 downstream of TDS2. For uniform bunch compression with $C^{-1}=(1+h R_{56})$, no acceleration in Linac2, i.e., $E_2=E_1$, and taking into account that $R_{33}=\pm\sqrt{\beta_{y}(s_2)/\beta_{y}(s_1)}$, Eq.~(\ref{eq:cond}) is identical to Eq.~(\ref{eq:case1}). Thus, both formalisms yield the same result.

		Since the kick strength of the first TDS is very weak, it can be implemented by means of a short rf structure and the thin-lens approximation is still valid. However, the kick strength of the second TDS is usually stronger, and the effect of the finite structure length should be taken into account. The symplectic beam transport matrix of a finite TDS with the length $L_2$ is given in Ref.~\cite{exchange1} by
		\begin{equation}
		{\mathbf R}_T^{thick}=\begin{pmatrix}
		1 & L_2 & K_2 L_2/2  & 0 \\
		0 & 1 & K_2 & 0 \\
		0 & 0 & 1 &0 \\
		K_2 & K_2L_2/2 & K_2^2L_2/6 & 1
		\end{pmatrix}.
		\label{eq:MTDS_thick}
		\end{equation} In this case, we require the point-to-point imaging is from the first TDS to the middle of the second TDS in order to have a complete cancellation. 
The overall matrix from TDS1 to the end of TDS2, when Eq.~(\ref{eq:cond}) is fulfilled, becomes more complicated.
A few correction terms containing the length $L_2$ of TDS2 appear, which however does not change the working principle of the reversible beam heater system. It should be pointed out that downstream of the reversible heater system, the beam is slightly coupled in $y'-\delta_0$ and $y-z_0$, which results in a small growth of the projected emittance given by
	\begin{equation}
	\epsilon_{y,z}^2
	=\epsilon_{y0,z0}^2 +\epsilon_{y0} \epsilon_{z0} \frac{\beta_{y0} \gamma_{z0} K_1^2 R_{56}^2} {(1+h R_{56})^2}\,,
	\label{eq:emi}
	\end{equation}
	where $\epsilon_{y0,z0}$ is the initial vertical (longitudinal) emittance, and $\beta_{y0}$ and $\gamma_{z0}$ are the initial Twiss parameters. As is shown in the following section, this projected emittance growth is typically negligible.

	\section{Reversible heating and emittance preservation}\label{sec:Heating}

\begin{figure}[htb]
	\centering
	\includegraphics[width=1\linewidth]{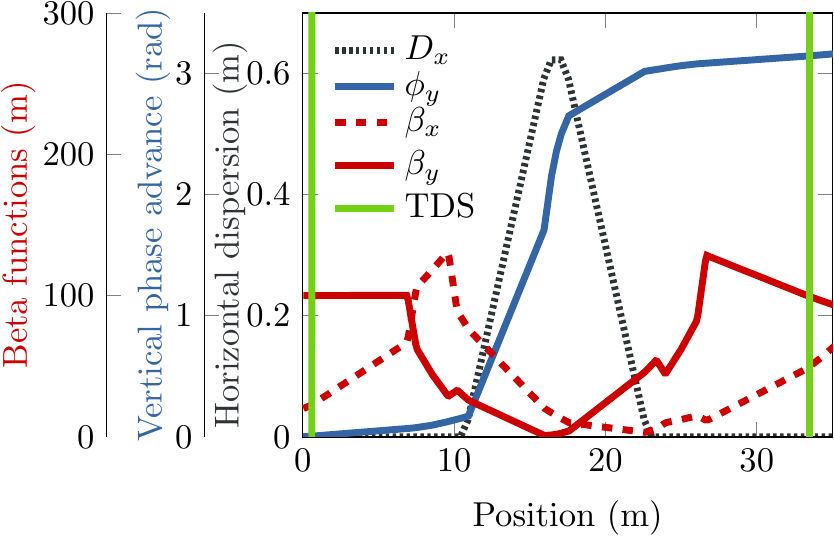}
	\caption{Relevant accelerator optics (Twiss parameters) and positions of the transverse deflecting rf structures used to numerically demonstrate the reversible beam heater system.} \label{fig:Fig2_Optics}
	\end{figure}

	We demonstrate the feasibility of the reversible beam heater system by numerical simulations using the particle tracking code {\it elegant}~\cite{Elegant}, and the simulations in the following include $5\,\times\,10^5$ particles. Table~\ref{tab:spec} summarizes the main parameters used in the simulations, and the adopted accelerator optics model, including the positions of the TDSs, is shown in Fig.~\ref{fig:Fig2_Optics}. The magnetic bunch compressor chicane is assumed to bend in the horizontal plane, and the TDSs are oriented perpendicularly with vertical deflection. In the previous section,
	\begin{table}[b]
	\centering
	\caption{Parameters of the electron beam, of the bunch compressor system, and of the transverse deflecting rf structures.}
	\begin{ruledtabular}
	\begin{tabular} {lcccc}
	Parameter           & Symbol       & Value     & Unit  \\ \hline
	Beam energy at TDS1/2    &$E$   & 350    & MeV \\
	Lorentz factor at TDS1/2    &$\gamma$   & 685    &  \\
	Initial transverse emittance & $\gamma \epsilon_{x,y}$ & 0.6 & $\mu$m \\
	Initial slice energy spread & $\sigma_E$ & $\sim$\,1  & keV \\
	Momentum compaction factor  &$R_{56}$  &$-138$      & mm \\
	Compression factor             &$C$      &$\sim$\,13  &      \\
	Final bunch current    &$I_f$    &$\sim$\,520  & A \\
	TDS1/2 rf frequency & $\omega/2\pi$ & 3.9 & GHz \\
	Voltage of TDS1 & $V_1$  & 0.415 & MV \\
	Voltage of TDS2 (without CSR)& $V_2$ &  5.440 & MV \\
	Length of TDS1 & $L_1$  & 0.1 & m \\
	Length of TDS2 & $L_2$ &  0.5& m \\
	\end{tabular}
	\label{tab:spec}
	\end{ruledtabular}
	\end{table}	
	 we included Linac2 for a general derivation of the method, but in practice, due to wakefield concerns, we recommend putting TDS2 right after the bunch compressor. In order to show numerical examples based on this approach, Linac2 is not considered anymore throughout the rest of this paper. Except for the TDSs, the parameters are similar to the magnetic bunch compressor system discussed for the Next Generation Light Source at LBNL~\cite{Corlett,Venturini2}.

	The initial longitudinal electron bunch  profile is assumed to be flat-top with a peak current of $\sim$\,40\,A and a slice energy spread of $\sim$\,1\,keV (r.m.s.). The initial linear and quadratic chirp is set for a uniform compression factor $C$ of about 13 across the entire bunch length. This is possible even with bunch compressor nonlinearities by using a higher-harmonic rf linearizer upstream of the bunch compressor~\cite{Lin} and needed to achieve uniform cancelation of the induced energy spread downstream of TDS2.

\begin{figure}[b]
	\centering
	\subfigure[~Upstream of TDS1.]{\includegraphics[width=0.48\linewidth]{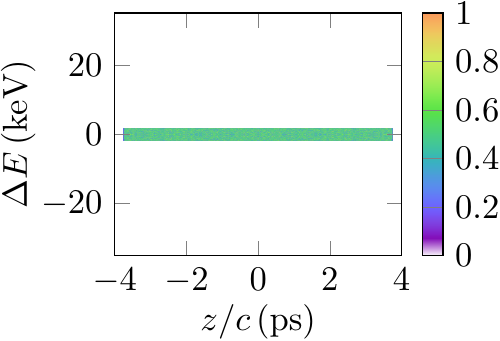} \label{fig:Fig3_Principle_1}}
	\subfigure[~Downstream of TDS1.]{\includegraphics[width=0.48\linewidth]{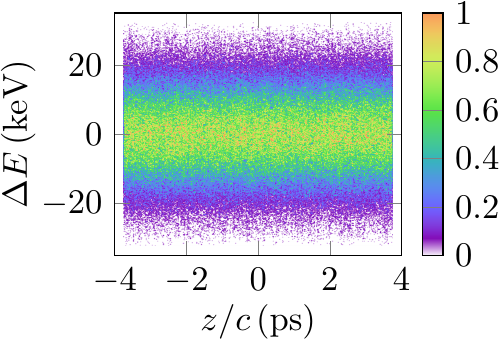} \label{fig:Fig3_Principle_2}}
	\subfigure[~Upstream of TDS2.]{\includegraphics[width=0.48\linewidth]{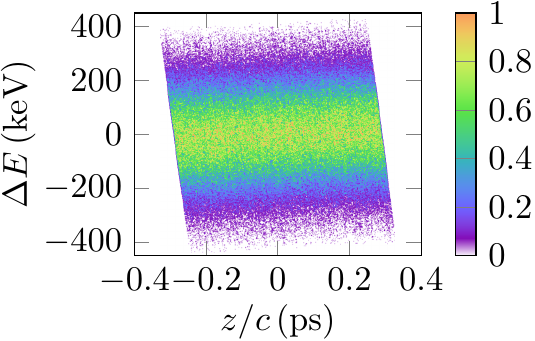} \label{fig:Fig3_Principle_3}}
	\subfigure[~Downstream of TDS2.]{\includegraphics[width=0.48\linewidth]{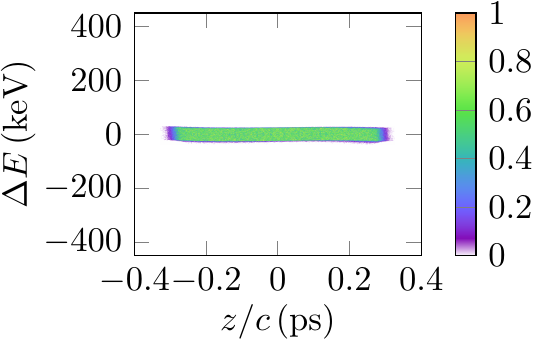} \label{fig:Fig3_Principle_4}}
	\caption{Simulation of the longitudinal phase space after removing the correlated energy chirp: (a) upstream of the first TDS, (b) directly downstream of the first TDS, (c) directly downstream of the bunch compressor and upstream of the second TDS, and (d) downstream of the second TDS. The axes scales change from (b) to (c) when bunch compression takes place. The bunch head is on the left, i.e., where $z/c<0$.} \label{fig:Fig3_Principle}
	\end{figure}

\begin{figure}[t]
	\centering
	\subfigure[~Projected emittances and core energy spread.]{\includegraphics[width=0.85\linewidth]{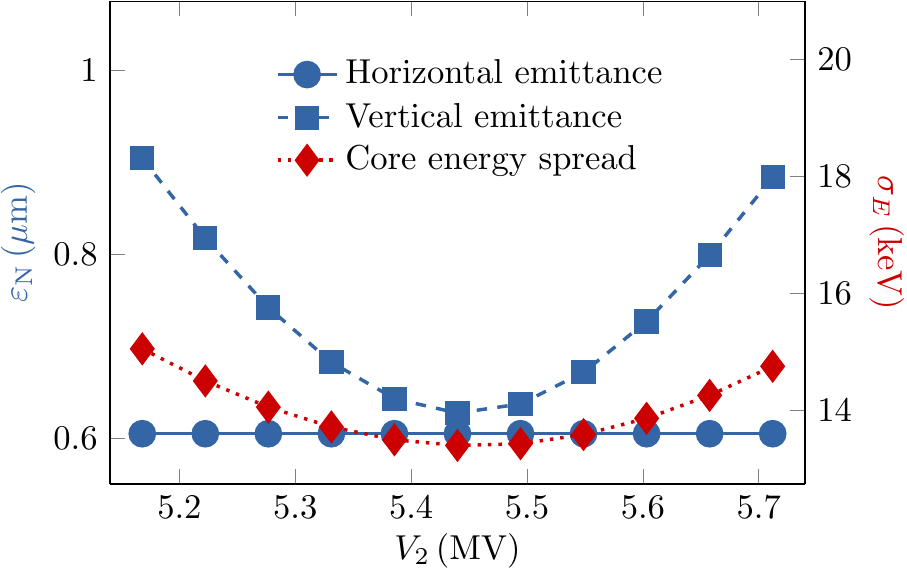} \label{fig:Fig4_a_CSR-off}}
	\subfigure[~Slice energy spread.]{\includegraphics[width=0.85\linewidth]{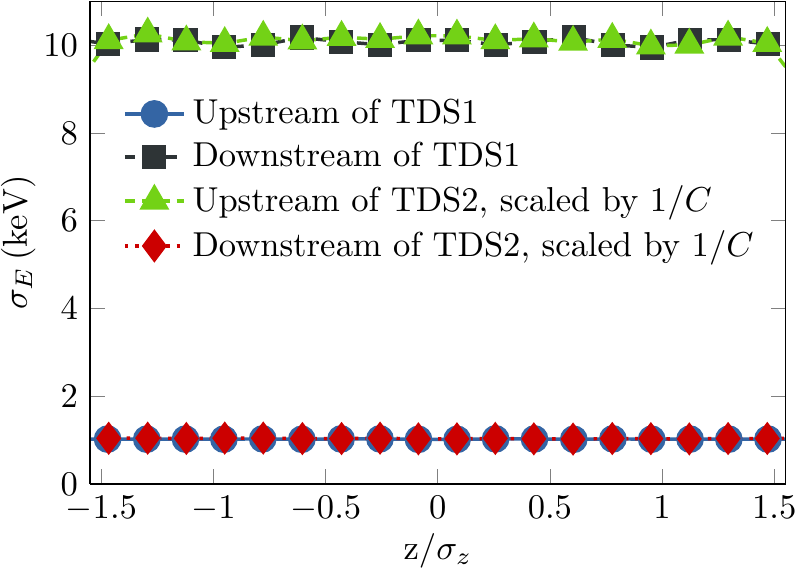} \label{fig:Fig4_b_CSR-off}}
	\caption{Simulations without CSR effects on the impact of the reversible heater system on projected emittances, core energy spread, and slice energy spread : (a) Projected emittances (normalized) and core energy spread, and (b) slice energy spread for $V_ {\mathrm{2}}$ at minimum emittance (see Fig.~\ref{fig:Fig4_a_CSR-off}). The longitudinal coordinate is normalized to the bunch length.} \label{fig:Fig4_CSR-off}
	\end{figure}

	Figure~\ref{fig:Fig3_Principle} shows the principle of the reversible beam heater system by means of simulation of the longitudinal phase space at different positions along the beamline.
	 The impact of CSR is not taken into account (cf. next subsection for CSR effects). The initial slice energy spread is heated up to $\sim$\,10\,keV (r.m.s.) in the first TDS, increased by the compression factor in the bunch compressor to $\sim$\,130\,keV (r.m.s.), and finally cooled down to $\sim$\,13\,keV (r.m.s.) by the second TDS (see Figs.~\ref{fig:Fig3_Principle_1}-\ref{fig:Fig3_Principle_4}). The plot in Fig.~\ref{fig:Fig4_b_CSR-off} shows that the heating induced by the first TDS is perfectly reversible, and the final slice energy spread is simply the initial slice energy spread scaled with the compression factor, which would be exactly the same like in the case without using the reversible beam heater system. Figure~\ref{fig:Fig4_a_CSR-off} shows the heater system impact on both the projected emittance (horizontal and vertical) and the core energy spread, i.e., the slice energy spread in the center of the bunch, for different voltages in the second TDS. The minimum of the vertical emittance is related to the cancelation of the spatial chirp and energy spread induced by the first TDS. The horizontal emittance is not affected at all, and the small projected emittance growth (6~\%) in the vertical plane at the minimum is due to residual coupling generated by the system that is described by Eq.~(\ref{eq:emi}). Nevertheless, as shown in Sec.~\ref{subsec:CSR}, even in the case with CSR effects, the horizontal slice emittance stays unaffected at all and the vertical slice emittance exhibits only deviations in the bunch head ($z/c<0$) and tail ($z/c>0$).

	\subsection{Impact of coherent synchrotron radiation}\label{subsec:CSR}
\begin{figure}[b]
	\centering
	\subfigure[~Projected emittances and core energy spread.]{\includegraphics[width=0.85\linewidth]{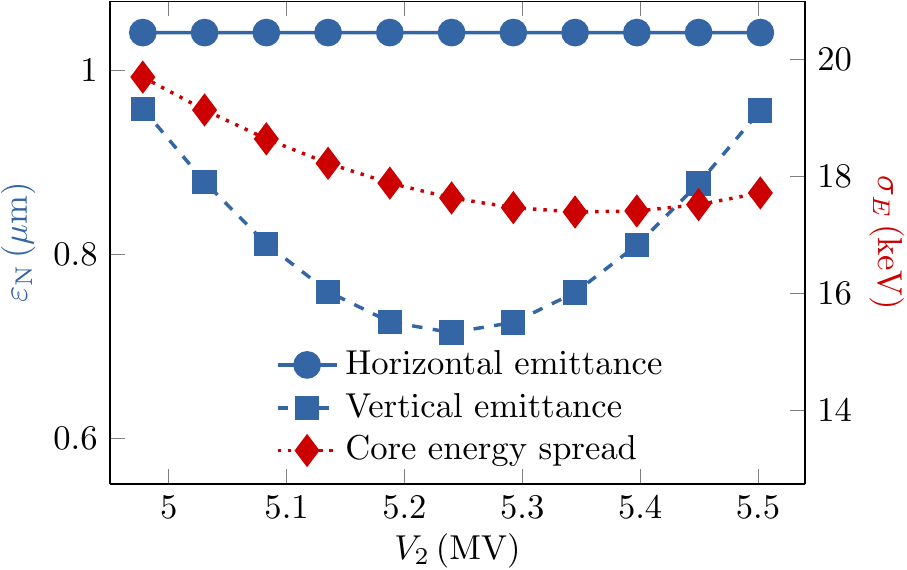} \label{fig:Fig5_a_CSR-on}}
	\subfigure[~Slice energy spread.]{\includegraphics[width=0.85\linewidth]{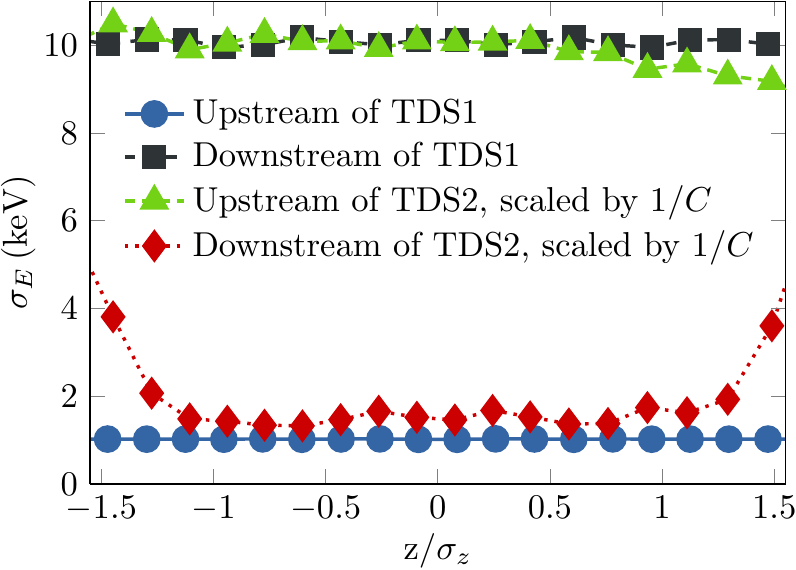} \label{fig:Fig5_b_CSR-on}}
	\caption{Simulation on the impact of the reversible beam heater system on projected emittances, core energy spread, and slice energy spread: (a) Projected emittances (normalized) and core energy spread, and (b) slice energy spread for $V_ {\mathrm{2}}$ at minimum emittance (see Fig.~\ref{fig:Fig5_a_CSR-on}). CSR effects are included by means of the 1-dimensional model in {\it elegant}~\cite{Elegant}.} \label{fig:Fig5_CSR-on}
	\end{figure}

\begin{figure}[t]
	\centering
	\includegraphics[width=0.85\linewidth]{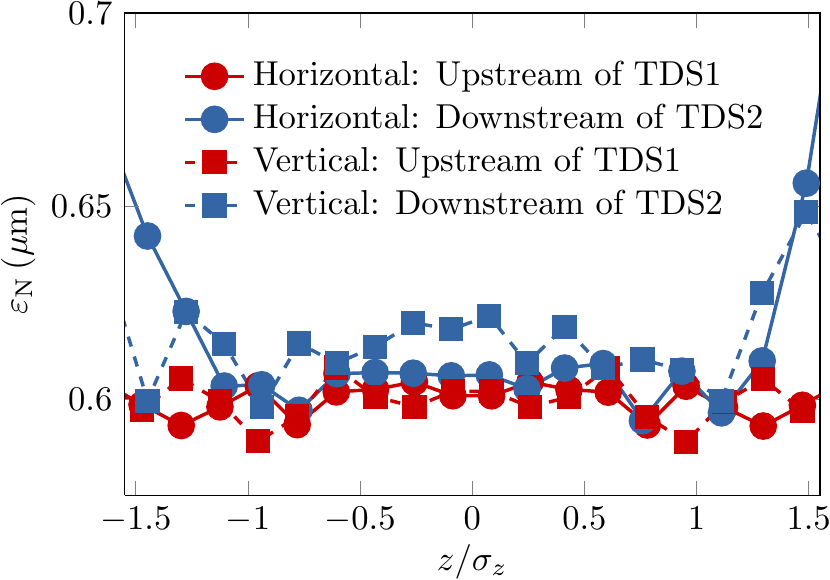}
	\caption{Simulation of the normalized slice emittance for both the vertical and horizontal upstream of the first and downstream of the second TDS. CSR effects are included.} \label{fig:Fig6_Emit}
	\end{figure}

	The previous results undergo small modifications when including CSR effects, which is shown in Fig.~\ref{fig:Fig5_CSR-on}. The voltage of the second TDS for minimum projected emittance in the vertical is shifted by about 0.2\,MV to lower values which is due the additional energy chirp induced by CSR. In comparison to the case without any CSR effects (cf. Fig.~\ref{fig:Fig4_CSR-off}), the projected emittance in the vertical plane is slightly increased and the slice energy spread is not perfectly canceled in the head and tail. The slice energy spread in the core part of the bunch is also slightly increased to 17.5\,keV (r.m.s.) (instead of 13.5\,keV (r.m.s.) in the absence of CSR). The projected emittance in the horizontal is about 1.7 larger which is independent of the reversible beam heater operation. This horizontal emittance growth can further be reduced by minimizing the horizontal beta function in the last dipole of the chicane where the bunch length becomes the shortest. This optimization is independent of the relevant motion in the vertical and does not affect the results of the reversible heater system.
	Albeit the fact that the projected emittances are increased, the horizontal slice emittance stays unaffected and the vertical slice emittance exhibits only deviations in the head and tail due to CSR effects as is shown in Fig.~\ref{fig:Fig6_Emit}. Thus, the core emittances are well preserved. We note that vertically streaked bunches in the bunch compressor chicane may change the impact of CSR effects but require a 3-dimensional ``point-to-point'' tracking which is not available neither in \mbox{{\it elegant}} nor in \mbox{{\it CSRtrack}~\cite{CSRtrack}}, and is beyond the scope of this paper.

	\section{Microbunching gain suppression}\label{sec:uB}
	The principle of the microbunching gain suppression with the reversible beam heater system is shown by an analytical treatment following Ref.~\cite{LCLSheater} and by using the beam transport matrix in Eq.~(\ref{eq:MTDS}). Then we show the feasibility of the reversible heater system to suppress microbunching instabilities by means of particle tracking simulations with initial density and energy modulations.
		\subsection{Analytical calculations}
		Using the vector notation $(y_0,y_0',z_0,\delta_0)$ for particles in the first linac upstream of the first TDS, the longitudinal position downstream of the second TDS is given by
		\begin{equation}
		z=K_1 R_{56} y_0 + (1+h R_{56}) z_0+R_{56} \delta_0\,.
		\end{equation}
		
		Suppose that $\delta_0=\delta_u+\delta_m$, where $\delta_u$ is the uncorrelated relative energy deviation, and $\delta_m(z_0)$ is the relative energy modulation accumulated before and in the first linac (Linac1). Following Ref.~\cite{LCLSheater}, the initial energy modulation at the wavenumber $k_0$ is converted into additional density modulation at a compressed wavenumber $k$. For a 4-dimensional (4-D) distribution function $F(y,y',z,\delta)$, the bunching factor $b(k)$ is given by
		\begin{align}
			b(k) =& \int dy dy' dz d\delta e^{-ik z} F(y,y',z,\delta) \nonumber \\
			=& \int dy_0 dy_0' dz_0 d\delta_u
			e^{-ik K_1 R_{56}y_0-ik (1+h R_{56}) z_0} \nonumber \\
			& e^{ -ik R_{56}(\delta_u+\delta_m(z_0))} F_0(y_0,y_0',z_0,\delta_u)
			\,,
		\label{eq:bunching}
		\end{align}
		where $F_0(y_0,y_0',z_0,\delta_u)$ is the initial 4-D distribution.
		If the induced energy modulation is small such that $\vert k
		R_{56} \delta_m \vert \ll 1$, we can expand the exponent of Eq.~(\ref{eq:bunching}) up to
		the first order in $\delta_m$ to obtain
		\begin{align}
			b(k) \approx &~b_0 (k_0) - i k R_{56} \int dz_0 \delta_m (z_0)  e^{-ik_0 z_0}\nonumber \\
		\times & \int dy_0 d\delta_u  e^{-ik K_1 R_{56}y_0-ik R_{56} \delta_u}  	
		U(y_0) V(\delta_u)   \,,
		\label{eq:bunching2}
		\end{align}
		where $k =C k_0$, $C=1/(1+h R_{56})$,  $U(y_0)$ describes the transverse profile, and $V(\delta_u)$ is the initial energy distribution. For both Gaussian profiles ($U$ and $V$), we have
		\begin{align}
			b(k) = b_0 (k_0) - & i k R_{56} \delta_m (k_0) \exp\left[-(k^2 R_{56}^2 K_1^2 \sigma_{y1}^2/2)\right]  \nonumber \\
		\times & \exp\left[-(k^2 R_{56}^2\sigma_{\delta u}^2/2)\right]\,.
		\label{eq:bunching3}
		\end{align}
		Here, we denote the Fourier transform of $\delta_m(z_0)$ as $\delta_m(k_0)$, which is the accumulated energy modulation at the wavenumber $k_0$ in the first linac due to longitudinal space charge and other collective effects. The initial energy spread is given by $\sigma_{\delta u}$, and $\sigma_{y1}$ is the vertical beam size in the first TDS. We see that $K_1 \sigma_{y1}$ acts like effective energy spread for microbunching gain suppression.

		\subsection{Numerical simulations}
		
		\begin{figure}[htb]
		\centering
		\subfigure[~Downstream of TDS2: Heater system off.]{\includegraphics[width=0.88\linewidth]{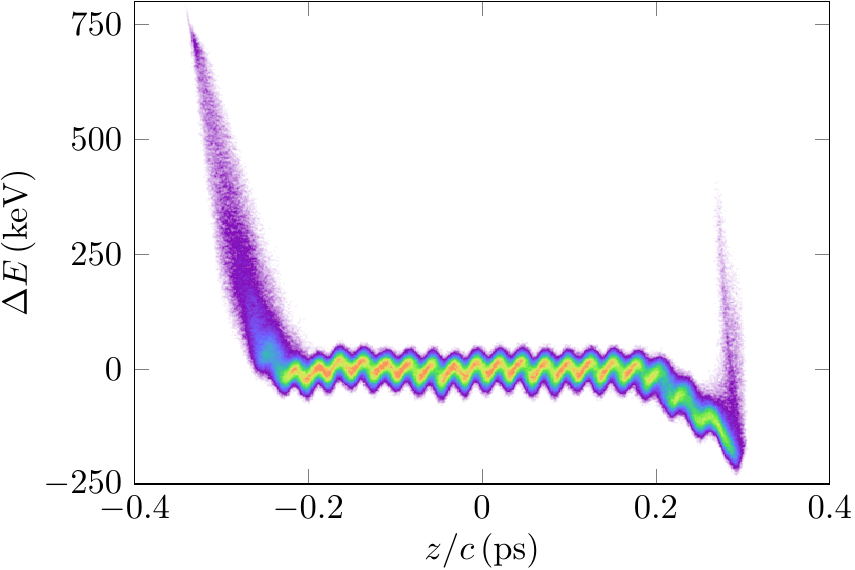} \label{fig:Fig7_Density_off}}
		\subfigure[~Downstream of TDS2: Heater system on.]{\includegraphics[width=0.88\linewidth]{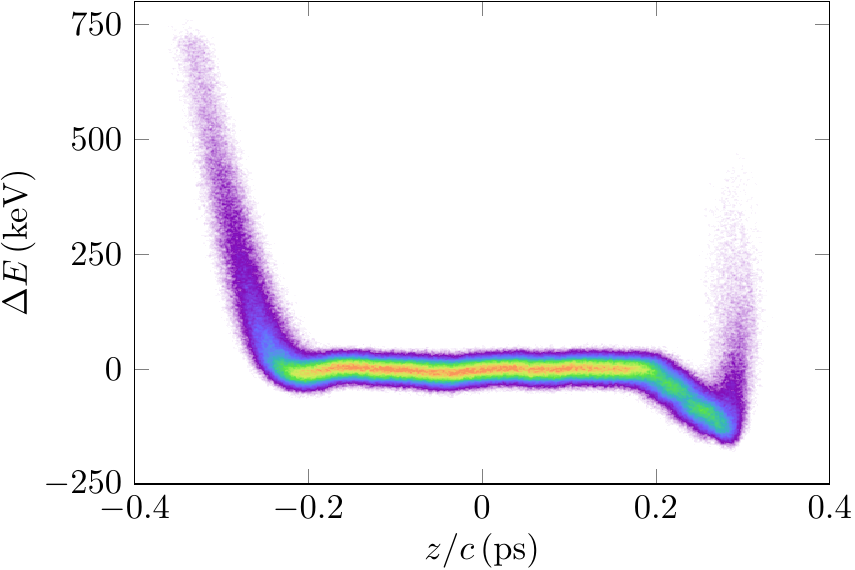} \label{fig:Fig7_Density_on}}
		\caption{Simulation on suppression of microbunching instabilities due to an initial density modulation, i.e., simulating CSR-driven microbunching. The entire longitudinal phase space, after removing the correlated energy chirp, is shown.} \label{fig:Fig7_Density}
		\end{figure}
\begin{figure}[htb]
		\centering
		\subfigure[~Downstream of TDS2: Heater system off.]{\includegraphics[width=0.88\linewidth]{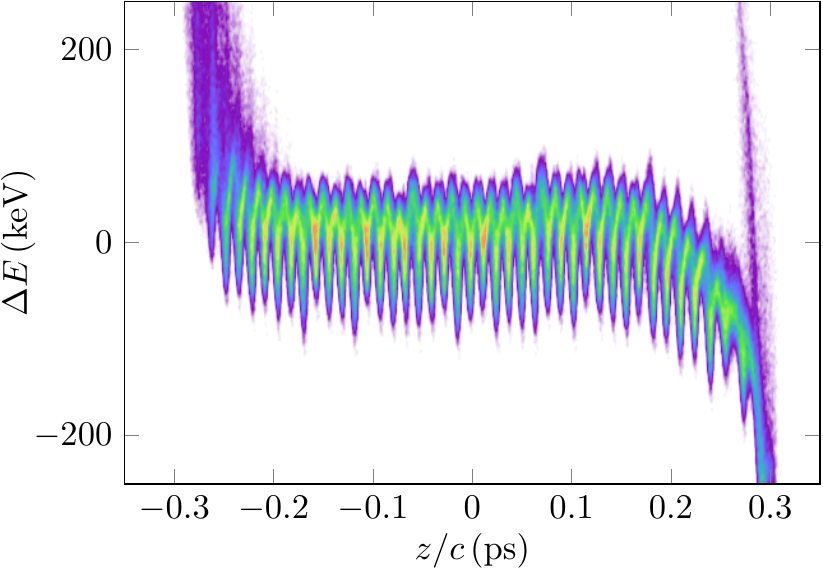} \label{fig:Fig8_Energy_off}}
		\subfigure[~Downstream of TDS2: Heater system on.]{\includegraphics[width=0.88\linewidth]{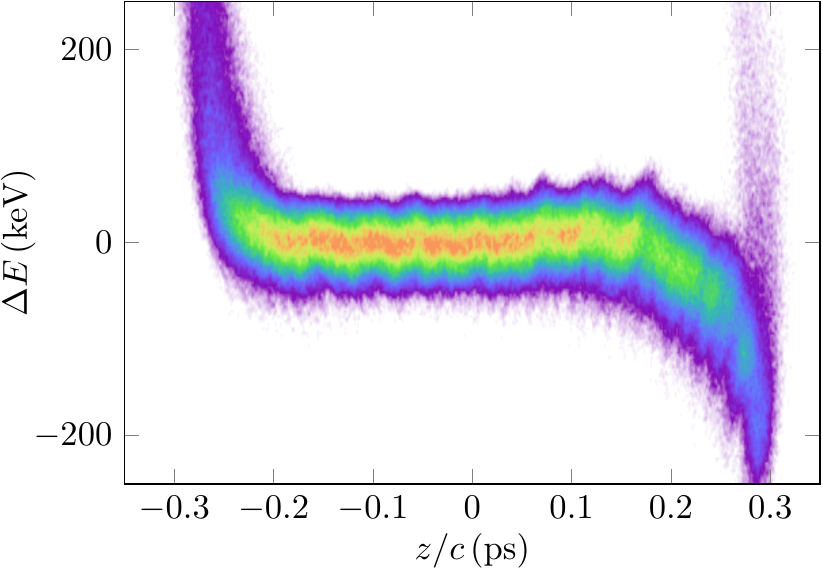} \label{fig:Fig8_Energy_on}}
		\caption{Simulation on suppression of microbunching instabilities due to an initial energy modulation, i.e., simulating longitudinal space charge driven microbunching. For the sake of clarity, only the core of the longitudinal phase space, after removing the correlated energy chirp, is shown.} \label{fig:Fig8_Energy}
		\end{figure}

		Suppression of microbunching instabilities is demonstrated by using both a pure initial density modulation with 5\,$\%$ peak amplitude and $100\,\mathrm{\mu m}$ modulation wavelength ($\lambda_m$), and a pure initial energy modulation with 3\,keV peak amplitude and $\lambda_m=50\,\mathrm{\mu m}$. Whereas the case with initial energy modulation is immediately consistent with the previous analytical treatment and describes the longitudinal space charge driven microbunching instability~\cite{lsc-ub}, the initial density modulations need to be converted into energy modulations by longitudinal CSR-impedance which expresses the consistency and describes the CSR-driven microbunching instability~\cite{CSR-ub}. The simulations were performed using the code \mbox{{\it elegant}} with $1\,\times\,10^6$ particles. Figure~\ref{fig:Fig7_Density} shows the longitudinal phase space downstream of the second TDS, after removing the correlated energy chirp (linear and quadratic chirp), for both the reversible beam heater system switched off (Fig.~\ref{fig:Fig7_Density_off}) and on (Fig.~\ref{fig:Fig7_Density_on}). In the case without reversible beam heater, energy and density modulations at the compressed modulation wavelength $\lambda_m/C$ appear, i.e., CSR-driven microbunching becomes visible. When switching the reversible beam heater on, the microbunching instability disappears and the resulting longitudinal phase space remains smooth. The reason is that the microbunches at the compressed wavelength are smeared due to $R_{56}K_1\sigma_{y1}$ (cf. Eq.~(\ref{eq:bunching3})), and accordingly, the modulations appear as correlations in the phase spaces $(y,z)$ and $(y',\delta)$.	The same effect of microbunching suppression is given for initial energy modulations as shown in Fig.~\ref{fig:Fig8_Energy}. 
		 The effect of the microbunching instability appears even stronger compared to the simulations case with initial density modulations, but the performance of the reversible heater system is the same with a smooth residual longitudinal phase space when the reversible beam heater is switched on (see Fig.~\ref{fig:Fig8_Energy_on}).

		Figures~\ref{fig:Fig7_Density} and~\ref{fig:Fig8_Energy} are obtained for a magnetic bunch compressor system as shown in Fig.~\ref{fig:Fig1_Setup}. The electron bunch will be further accelerated and transported throughout the rest of the accelerator to reach the final beam energy and peak current in order to drive an X-ray FEL (not studied in this paper). A microbunched electron beam as illustrated in Figs.~\ref{fig:Fig7_Density_off} and~\ref{fig:Fig8_Energy_off}, i.e., when the reversible beam heater system is switched off, will accumulate additional energy and density modulations, which would lead to unacceptable longitudinal phase space properties for an X-ray FEL such as a large slice energy spread.

	\section{Practical considerations}\label{sec:Practical}
	The previous sections covered the principle of reversible electron beam heating and microbunching gain suppression by means of analytical calculations and numerical simulations. In real accelerators, we also have to deal with imperfections, jitter and drifts of various parameters, and accordingly supplementary studies with respect to sensitivity on jitter sources and tolerances have to be performed. In the following, we discuss the impact of beam and rf jitter on the reversible beam heater system, and also point out the inherent possibility of longitudinal phase space diagnostics and on-line monitoring.

		\subsection{Jitter and tolerances}
		The impact of beam and rf jitter on the reversible beam heater method can effectively be discussed using the Eqs.~(\ref{eq:long}) and~(\ref{eq:MTDS}) with the condition in Eq.~(\ref{eq:cond}). Deviations from the conditions in Eq.~(\ref{eq:cond}) can appear from jitter of the individual peak deflection voltages $V_1$ and $V_2$ of the TDSs, and lead to growth of the projected vertical emittance as is shown in Fig.~\ref{fig:Fig5_a_CSR-on}, where the voltage of the second TDS is varied. Even in the case of a large TDS voltage jitter of 1\,\%, the vertical projected emittance growth is less than 2\,$\%$ (see Fig.~\ref{fig:Fig5_a_CSR-on}). In the case of acceleration between the first and second TDS, also energy jitter, which is similar or smaller than TDS voltage jitter, due to this intermediate acceleration leads to deviation of the condition in Eq.~(\ref{eq:cond}). The choice of superconducting accelerator technology even provide much better rf stability~\cite{HS,CS}. Pure arrival time jitter upstream of the first TDS has no impact as long as the condition in Eq.~(\ref{eq:cond}), which describes the coupling between $y'$ and $t=z/c$, is fulfilled. In the case that Eq.~(\ref{eq:cond}) is not exactly fulfilled, e.g., due to TDS voltage jitter which is on the percent-level, the impact of typical arrival time jitter well below 100\,fs, like at the LCLS~\cite{LCLSnature2} or FLASH~\cite{CS}, is negligible. The most critical jitter sources arise from energy jitter upstream of the bunch compressor chicane and from rf phase jitter in the TDSs. The momentum compaction factor translates energy jitter into arrival time jitter, which leads to vertical kicks in the second TDS. The same effect of additional vertical kicks is generated by rf phase jitter in the TDSs. In order to have small impact of vertical kicks on the remaining beam transport, we demand $\Delta \sigma_{y'}\ll\sigma_{y'}$ directly downstream of the second TDS with the induced vertical r.m.s.~kick $\Delta\sigma_{y'}$ and the intrinsic beam divergence $\sigma_{y'}$. The relevant total vertical r.m.s.~kick is given by
		\begin{align}
		\Delta \sigma_{y'} =& \sqrt{\left(K_2 R_{56}\frac{\sigma_{E}}{E} \right )^2+\left(K_2\frac{c}{\omega}\sigma_{\varphi_2} \right )^2 +\left (\frac{K_1}{R_{33}}\frac{c}{\omega}\sigma_{\varphi_1} \right )^2} \nonumber \\
				= &\sqrt{\left(K_2 R_{56}\frac{\sigma_{E}}{E} \right )^2+\left(K_2\frac{c}{\omega}\right )^2\left(\sigma_{\varphi_2}^2 + \frac{1}{C^{2}}\sigma_{\varphi_1}^2 \right ) }\nonumber \\
				\approx& \sqrt{\left(K_2 R_{56}\frac{\sigma_{E}}{E} \right )^2+\left(K_2\frac{c}{\omega}\right )^2 \sigma_{\varphi_2}^2 }
		\end{align}
		with the energy jitter $\sigma_{E}/E$ upstream of the bunch compressor, the rf phase jitter $\sigma_{\varphi_{1,2}}$ of the TDSs, the magnification factor $R_{33}$ from the first to the second TDS (see Eq.~(\ref{eq:-I})), and using Eq.~(\ref{eq:cond}) with the compression factor $C=(1+hR_{56})^{-1}$. We see that the vertical r.m.s.~kick due to rf phase jitter in the first TDS scales with $C^{-2}$ and can be neglected compared to the vertical r.m.s.~kick induced by the second TDS when we assume the same amount of rf phase jitter in both TDSs. The condition for trajectory stability $\Delta \sigma_{y'}\ll\sigma_{y'_2}=\sqrt{\epsilon_{y_2}/\beta_{y_2}}$ with the intrinsic (uncorrelated) r.m.s.~beam divergence $\sigma_{y'_2}$ downstream of the bunch compressor at TDS2, where $\epsilon_{y_2}$ is the geometrical emittance, can be restated as 
		\begin{equation}
		\sqrt{\left(R_{56}\frac{\sigma_{E}}{E}\right)^2+\left(\frac{c}{\omega}\sigma_{\varphi_2} \right )^2} \ll \frac{\epsilon_{y_2}}{K_2 \sqrt{\beta_{y_2} \epsilon_{y_2}}}=\frac{\sqrt{\epsilon_{y_2}\epsilon_{y_1}}}{C \Delta \sigma_{\delta_1}}\,.
		\label{eq:stab}
		\end{equation} Here, $\Delta \sigma_{\delta_1}$ is the additional relative energy spread induced by the first TDS for suppression of microbunching instabilities, and $\epsilon_{y_1}$ denotes the geometrical emittance upstream of the bunch compressor at TDS1.

		For the example parameters discussed throughout this paper (see also Table~\ref{tab:spec}), i.e., $C=13$, $\gamma\epsilon_{y_1}=0.6\,\mathrm{\mu m}$, $\gamma\epsilon_{y_2}=0.72\,\mathrm{\mu m}$ (see Fig.~\ref{fig:Fig5_a_CSR-on}), and $\Delta \sigma_{\delta_1}E\approx10\,\mathrm{keV}$ with $E=350\,\mathrm{MeV}$ ($\gamma=685$), the stability condition in Eq.~(\ref{eq:stab}) yields pure relative energy jitter (neglecting rf phase jitter) of $\sigma_{E}/E\ll1.9\cdot10^{-5}$ or pure rf phase jitter (neglecting energy jitter) of $\sigma_{\varphi_2}\ll0.012^\circ$. A combination of both will obviously tighten the acceptable jitter. This level of rf stability is difficult to achieve in normal conducting linacs with single bunch operation, but might be achieved with superconducting accelerator technology like at FLASH or as planned for NGLS, where many bunches can be accelerated in a long rf pulse, i.e., in a bunch train. Currently, several rf feedforward and feedback controls are able to stabilize the bunches at FLASH to $\sigma_{E}/E=3.0\cdot10^{-5}$ and $\sigma_{\varphi}=0.007^\circ$ at 150\,MeV~\cite{HS,SP}, and further improvements towards $\sigma_{E}/E\leq1.0\cdot10^{-5}$ are planned using a fast normal conducting cavity upstream of the bunch compressors~\cite{CS,HS}. With perfect scaling of rf jitter from several independent rf power stations that adds uncorrelated, we would expect an improvement of $\sqrt{150\,\mathrm{MeV}/350\,\mathrm{MeV}}\approx 0.66$ compared to the results at FLASH with $150\,\mathrm{MeV}$ and assuming the beam energy of $350\,\mathrm{MeV}$ in the bunch compressor of the NGLS design. Continuous-wave rf operation, as planned for the NGLS design~\cite{Corlett}, and a proper choice of rf working points for FEL operation might improve the stability further.

		\subsection{Integrated longitudinal phase space diagnostics}
		A practical spin-off of the reversible beam heater system is the availability of longitudinal phase space diagnostics. The vertical betatron motion of electrons passing through a TDS is described by Eq.~(\ref{eq:motion}), which enables a mapping from time (longitudinal coordinate) to the vertical~\cite{Kick,Roehrs,Filippetto}, and finally a possibility to obtain temporal bunch information by means of transverse beam diagnostics. In a similar manner, the relative energy deviation is mapped to the horizontal in the presence of horizontal momentum dispersion, like in a magnetic bunch compressor chicane (see, e.g., Refs.~\cite{Roehrs,Filippetto}). The combined operation makes single-shot measurements of the longitudinal phase space possible, and in the case of the generic layout of a reversible electron beam heater system as depicted in Fig.~\ref{fig:Fig1_Setup}, longitudinal phase space measurements become feasible using the first TDS and observation screens in the dispersive section of the bunch compressor chicane. In order to get information of the bunch length after the bunch compression, the second TDS can be used with downstream observation screens (not shown in Fig.~\ref{fig:Fig1_Setup}).
	
		In addition to invasive longitudinal phase space measurements of a single bunch using observation screens, even fully noninvasive measurements utilizing incoherent synchrotron radiation, emitted in the bunch compressor bending magnets, are possible (see, e.g., Ref.~\cite{Gerth}). When using a fast gated camera, the implication will be the possibility of on-line monitoring the longitudinal phase space of individual bunches in multi-bunch accelerators.

	\section{Summary and conclusions}\label{sec:Summary}
	Our studies show that the reversible beam heater system proposed here can suppress microbunching instabilities and preserve the high beam brightness at the same time. Due to CSR effects, some vertical emittance degradation in the head and tail region of the bunch occurs, but the core emittances are well preserved. In the numerical demonstrations using the code \mbox{{\it elegant}}, the first TDS generates about 10\,keV (r.m.s.) slice energy spread, which is similar to the laser heater but with a more Gaussian energy distribution (cf. laser heater). The bunch compression process increases the slice energy spread to $\sim$\,130\,keV (r.m.s.), which is then reversed to $\sim$\,17\,keV (r.m.s.) after the second TDS in the presence of CSR effects. Without CSR effects, the slice energy spread is reversed to $\sim$\,13\,keV (r.m.s.), which demonstrates perfect cancelation. The simulations also show that initial bunching in energy and density in the beam can be smeared out during the process in the reversible beam heater system, i.e., microbunching instabilities can be suppressed. The resulting smooth beam can then propagate through the remaining accelerator without further generation of much additional energy spread and is advantageous for any kind of laser seeding manipulations and experiments. For example, this scheme significantly loosen the required laser power for short-wavelength HHG seeding~\cite{NLS} and may strongly impact the design of future seeded FELs. In addition, the reversible beam heater system exhibits integrated options for diagnosis and on-line monitoring of the longitudinal phase space applicable for multi-bunch machines, which is also the preferred type of accelerator for the reversible heater system due to large sensitivities on energy and rf jitter. Linear accelerators based on superconducting rf technology might be able to match the strict tolerances in order to keep vertical kicks small and to achieve a sufficient trajectory stability in the downstream undulators.

	\begin{acknowledgments}
	We would like to thank P. Emma, Ch.~Gerth, A. Lumpkin, H. Schlarb, and J. Thangaraj for useful discussions and suggestions. This work was supported by Department of Energy Contract No. DE-AC02-76SF00515.

	\end{acknowledgments}


\begin{thebibliography}{38}

		\bibitem{LCLSnature1} S. Jamison, Nature Photonics  {\bf 4}, 589 - 591 (2010).

		\bibitem{FLASHAnature} W. Ackermann {\it et al.}, Nature Photonics {\bf 1}, 336 - 342 (2007).

		\bibitem{LCLSnature2} P. Emma {\it et al.}, Nature Photonics {\bf 4}, 641 - 647 (2010).
		
		\bibitem{SACLAnature} D. Pile, Nature Photonics {\bf 5}, 456 - 457 (2011).
	
		\bibitem{CSR} E.L. Saldin, E.A. Schneidmiller, and M.V. Yurkov, Nucl. Instrum. Methods Phys. Res., Sect. A {\bf 398}, 373 (1997).

		\bibitem{CSR-ub} M. Borland  {\it et al.}, Nucl. Instrum. Methods Phys. Res., Sect. A  {\bf 483}, 268 (2002).

		\bibitem{lsc-ub} E.L. Saldin, E.A. Schneidmiller, and M.V. Yurkov, Nucl. Instrum. Methods Phys. Res., Sect. A  {\bf 528}, 355 (2004).

		\bibitem{LH} Z. Huang, M. Borland, P. Emma, J. Wu, C. Limborg, G. Stupakov, and J. Welch, Phys. Rev. ST Accel. Beams {\bf 7}, 074401 (2004).

		\bibitem{LCLSheater} Z. Huang {\it et al.}, Phys. Rev. ST Accel. Beams {\bf 13}, 020703 (2010).

		\bibitem{NLS} D.J. Dunning {\it et al.}, Proceedings of the 1st International Particle Accelerator Conference, Kyoto, Japan, 2010, TUPE049.

        \bibitem{flatflat} M. Cornacchia, S. Di Mitri, G. Penco, and A. Zholents, Phys. Rev. ST Accel. Beams {\bf 9}, 120701 (2006).

		\bibitem{okly} Y. Ding, P. Emma, Z. Huang, and V. Kumar, Phys. Rev. ST Accel. Beams {\bf 9}, 070702 (2006).

		\bibitem{LOLA} O. Altenmueller, R. Larsen, and G. Loew, Rev. Sci. Instrum. {\bf 35}, 438 (1964).

		\bibitem{Kick} P. Emma, J. Frisch, and P. Krejcik, Technical Report No. LCLS-TN-00-12, 2000.
		
		\bibitem{Roehrs} M. R\"{o}hrs, Ch. Gerth, H. Schlarb, B. Schmidt, and P. Schm\"{u}ser, Phys. Rev. ST Accel. Beams {\bf 12}, 050704 (2009).

		\bibitem{Filippetto} D. Filippetto {\it et al.}, Phys. Rev. ST Accel. Beams {\bf 14}, 092804 (2011).

		\bibitem{xtcav} Y. Ding, C. Behrens, P. Emma, J. Frisch, Z. Huang, H. Loos, P. Krejcik, and M-H. Wang, Phys. Rev. ST Accel. Beams {\bf 14}, 120701 (2011).

		\bibitem{psxray} A. Zholents, P. Heimann, M. Zolotorev, and J. Byrd, Nucl. Instrum. Methods Phys. Res., Sect. A  {\bf 425}, 385 (1999).

        \bibitem{exchange1} M. Cornacchia and P. Emma, Phys. Rev. ST Accel. Beams {\bf 5}, 084001 (2002).

        \bibitem{exchange2} P. Emma, Z. Huang, K.-J. Kim, and P. Piot, Phys. Rev. ST Accel. Beams {\bf 9}, 100702 (2006).

        \bibitem{mapping} D. Xiang and Y. Ding, Phys. Rev. ST Accel. Beams {\bf 13}, 094001 (2010).
		
		\bibitem{ramp}	P. Piot, Y.-E Sun, J.G. Power, and M. Rihaoui1, Phys. Rev. ST Accel. Beams {\bf 14}, 022801 (2011).

        \bibitem{psex} D. Xiang and A. Chao, Phys. Rev. ST Accel. Beams {\bf 14}, 114001 (2011).

		\bibitem{Xiang} D. Xiang {\it et al.}, Phys. Rev. Lett. {\bf 105}, 114801 (2010).
		
		\bibitem{Xiang2} D. Xiang {\it et al.}, Phys. Rev. Lett. {\bf 108}, 024802 (2012).

		\bibitem{Corlett} J. Corlett {\it et al.}, Proceedings of the 24th Particle Accelerator Conference, New York, USA, 2011, TUOCS5.

		\bibitem{Venturini} M. Venturini and A. Zholents, Nucl. Instrum. Methods Phys. Res., Sect. A {\bf 593}, 53 (2008).

		\bibitem{Lin} H. Edwards, C. Behrens, and  E. Harms, Proceedings of the 25th International Linear Accelerator Conference, Tsukuba, Japan, 2010, MO304.
	
		\bibitem{Panowsky} W. Panofsky and W. Wenzel, Rev. Sci. Instrum. {\bf 27}, 967 (1956).
	
		\bibitem{Bro} M.J. Browman, Proceedings of the 15th Particle Accelerator Conference, Washington, D.C., USA, 1993, p. 800.
		
		\bibitem{IES1} S. Korepanov, M. Krasilnikov, F. Stephan, D. Alesini, and L. Ficcadenti, Proceedings of the 8th European Workshop on Beam Diagnostics and Instrumentation for Particle Accelerators, Venice, Italy, 2007, TUPB32.

		\bibitem{IES2} C. Behrens and Ch. Gerth, Proceedings of the 9th European Workshop on Beam Diagnostics and Instrumentation for Particle Accelerators, Basel, Switzerland, 2009, TUPB44.

		\bibitem{IES3} C. Behrens and Ch. Gerth, Proceedings of the 10th European Workshop on Beam Diagnostics and Instrumentation for Particle Accelerators, Hamburg, Germany, 2011, TUPD31.

		\bibitem{emma} P. Emma (private communication).

		\bibitem{Elegant} M. Borland, ANL/APS Report No. LS-287, 2000.

		\bibitem{Venturini2} M. Venturini {\it et al.}, Proceedings of the 24th Particle Accelerator Conference, New York, USA, 2011, THP180.

		\bibitem{CSRtrack} M. Dohlus and T. Limberg, Proceedings of the 26th International Free Electron Laser Conference, Trieste, Italy, 2004, MOCOS05.

		\bibitem{HS} H. Schlarb (private communication).

		\bibitem{CS} C. Schmidt {\it et al.}, Proceedings of the 33rd International Free Electron Laser Conference, Shanghai, China, 2011, THPA26.

		\bibitem{SP} S. Pfeiffer {\it et al.}, LLRF-2011 workshop, Hamburg, Germany, 2011, ``Feedback Strategies for longitudinal Beam Stabilization''.

		\bibitem{Gerth} Ch. Gerth, Proceedings of the 8th European Workshop on Beam Diagnostics and Instrumentation for Particle Accelerators, Venice, Italy, 2007, TUPC03.
		
	\end{thebibliography}
\end{document}